\documentclass[12pt,preprintnumbers,showpacs,amsmath,amssymb,prd,floatfix, nofootinbib]{iopart}

\usepackage{graphicx}
\usepackage{epsfig}
\usepackage{bm}
\usepackage{amsfonts}
\usepackage{caption}

\def\ol{\Omega_d}

\begin{document}

\title[]{Holographic dark energy with time varying ${n^2}$ parameter in non-flat universe}

\author{Bushra Majeed$^1$, Mubasher Jamil$^2$ and Azad A. Siddiqui$^3$}

\address{School of Natural Sciences (SNS), National University of Sciences and Technology
(NUST), Islamabad, Pakistan.}

\ead{bushra$\_$majeed18@yahoo.com$^1$, mjamil@sns.nust.edu.pk$^2$,
azad@sns.nust.edu.pk$^3$}{\bf Abstract}.We consider a holographic
dark energy model, with a varying parameter, $n$, which evolves
slowly with time. We obtain the differential equation describing
evolution of the dark energy density parameter, $\Omega_d$, for the
flat and non-flat FRW universes. The equation of state parameter in
this generalized
version of holographic dark energy depends on $n$.\\
\\PACS {98.80.-k, 95.36.+x}

\section{Introduction}
The current observations \cite{c1}-\cite{c4} strongly support that
our universe is in the accelerated expansion phase. In the standard
cosmological structure, the existence of a component with
antigravity effect is necessary for explaining this accelerated
expansion. Using different approaches, many models have been
suggested to explain the dark energy. One of them is the use of the
Einstein cosmological constant, but it suffers from two problems
(the ``fine-tuning'' and the ``coincidence'') \cite{c7}. The
dynamical dark energy models with a variable equation of state have
been investigated, scalar field models are one of these
\cite{essence}. Another interesting approach for exploring the
behavior of dark energy is through the use of the principles of
quantum gravitation \cite{Witten:2000zk}. Proposal of holographic
dark energy (HDE) is an example of such models
\cite{Cohen:1998zx}-\cite{Li:2004rb}. The expression for energy
density of HDE is:
\begin{equation}\label{de}
\rho_d=\frac{3n^2}{8\pi G L^2},
\end{equation}
where $L$ is the infrared (IR) cut-off, $n$ is a constant, and $G$
is the Newton gravitational constant. The holographic dark energy
scenario is one of the most widely studied model and there are many
versions of this model in the literature \cite{j12}-\cite{obs3}.

If we take $L$ as a Hubble horizon, it results in a wrong equation
of state (EoS) and the accelerated expansion of the universe can not
be obtained. However, this issue can be resolved in the case of
interacting HDE. This model does not work with the particle horizon
as well, but when $L$ is chosen as future event horizon, the results
favor the accelerated expansion.

There is no strong evidence for $n$ to be taken a constant, so the
HDE parameter, $n^2$, has a vital role in characterizing the
properties of the model. For example in future, the model could be
like a phantom or quintessence dark energy model depends on whether
the value of $n^2$ is larger or smaller than $1$ respectively. Model
of HDE with variable $n^2$ parameter for a flat universe has been
studied in \cite{CPL}. There are many models of HDE in literature
with a variable gravitational constant, $G$. Following the approach
adopted by Jamil et al. \cite{G}, here we investigate the HDE model
with a variable $n^2$ parameter, allowing the consequent
modifications to the EoS parameter, $w$, of the dark energy. Plan of
our work is as follows: In section \ref{model} we build the HDE
model with a time varying $n^2(z)$ and extract the evolution
equations for the dark energy density parameter. In section
\ref{discussion} we calculate the corrections to the parameter, $w$,
at low redshifts. In section \ref{Numerical Results} we demonstrate
the numerical results and in section \ref{Conclusions} we summarize
our results.

\section{Holographic Dark Energy (HDE) with variable ${n^2}$ parameter}
\label{model}

\subsection{Flat FRW geometry}

To construct the HDE model with a variable $n^2$, we consider the
flat Robertson-Walker geometry given as
\begin{equation}\label{met}
ds^{2}=-dt^{2}+a^{2}(t)\left(dr^{2}+r^{2}(d\theta^{2}+\sin^{2}\theta
d\varphi^{2})\right),
\end{equation}
where $a(t)$ is the scale factor and $t$ is the comoving time. The
first Friedmann equation is given by
\begin{equation}\label{FR1}
H^2=\frac{8\pi G}{3}(\rho_m+\rho_d),
\end{equation}
where $H$ is the Hubble parameter, $\rho_m=\frac{\rho_{m0}}{a^3}$
denotes the matter density, and $\rho_d$ is the dark energy density.
The present value of a quantity is represented by index $``0''$.
Using the density parameter, $ \Omega_d\equiv\frac{8\pi
G}{3H^2}\rho_d$, with Eq. (\ref{de}) we get
\begin{equation}
 \label{OmegaL2}
\ol=\frac{n^2(z)}{H^2L^2}.
\end{equation}
As discussed before, for flat universe, defining $L$ as the future
event horizon is the best option \cite{Hsu:2004ri,
Li:2004rb,Guberina,Guberina0}, i.e. taking $L\equiv R_ h(a)$ as
\begin{equation}
 R_ h(a)=a\int_t^\infty{dt\over
a(t)}=a\int_a^\infty{da\over Ha^2}~.\label{eh}
\end{equation}
To denote the time derivative we use a dot, and prime is used for
the differentiation with respect to the independent variable $\ln
a$, i.e. we acquire $\dot{J}=J'H$, for every quantity $J$.
Differentiating Eq. (\ref{OmegaL2}), using Eq. (\ref{eh}), and
$\dot{R}_h=HR_h-1$, we attain
\begin{equation}\label{OmegaLdif}
\frac{\Omega'_d}{\Omega^2_d}=\frac{2}{\Omega_d}\Big[\frac{n^\prime}{n}-1-\frac{\dot{H}}{H^2}+\frac{\sqrt{\Omega_d}}{n}\Big].
\end{equation}
We can see that the varying behavior of $n^2$ has become apparent.
To eliminate $\dot{H}$ we differentiate the Friedman equation
(\ref{FR1}) and use the expression
\begin{equation}\label{rhoodot}
\rho'_d=\rho_d\left(\frac{n'}{n}-2+\frac{2\sqrt{\Omega_d}}{n}\right),
\end{equation}
to get
\begin{equation}\label{Hdot}
2\frac{\dot{H}}{H^2}=-3+\Omega_d\left(1+\frac{2\sqrt{\Omega_d}}{n}\right)+\frac{2n'}{n}\Omega_d,
\end{equation}
where $n$ is considered to be time dependent. Finally, using Eq.
(\ref{Hdot}) in Eq. (\ref{OmegaLdif}) we have
\begin{equation}\label{OmegaLdif3}
\Omega_d'=\Omega_d(1-\Omega_d)\Big[1+\frac{2\sqrt{\Omega_d}}{n}\Big]+2\Omega_d(1-\Omega_d)\frac{n'}{n}.
\end{equation}
Note that the second term is the correction term appearing because
of variable $n$, here $n'/n$ is a dimensionless number.

\subsection{Non-flat FRW geometry}
Now we extend the work presented in the previous subsection for the
FRW universe with metric
\begin{equation}\label{metr}
ds^{2}=-dt^{2}+a^{2}(t)\left(\frac{dr^2}{1-kr^2}+r^2(d\theta^{2}+\sin^{2}\theta
d\varphi^{2})\right) ,\end{equation} where $(t,r,\theta,\varphi)$
are comoving coordinates and $k$ represents the spacial curvature
with $k = -1, 0, 1$ respectively corresponding to the open, flat and
the closed universes. In this geometry the first Friedmann equation
becomes
\begin{equation}\label{FR1nf}
H^2+\frac{k}{a^2}=\frac{8\pi G}{3}(\rho_m+\rho_d).
\end{equation}
In non-flat metric, the cosmological length, $L$, for the HDE model
takes the following form \cite{nonflat}
\begin{equation}\label{Lnonflat}
L\equiv\frac{a(t)}{\sqrt{|k|}}\,\mbox{sinn}\left(\frac{\sqrt{|k|}R_h}{a(t)}\right),
\end{equation}
with
\begin{equation}\frac{1}{\sqrt{|k|}}\mbox{sinn}(\sqrt{|k|}y)=\cases{
 \sin y & \, \, $k=+1$,\\
             y & \, \, $k=0$,\\
             \sinh y & \, \, $k=-1$.\\
}\end{equation} It is easy to find that
\begin{equation}\label{Ldot}
\dot{L}=H L-\mbox{cosn}\left(\frac{\sqrt{|k|}R_h}{a}\right),
\end{equation}
where
\begin{equation}\mbox{cosn}(\sqrt{|k|}y)=\cases{
 \cos y  & \, \, $k=+1$,\\
             1 & \, \, $k=0$,\\
             \cosh y & \, \, $k=-1$.\\
}\end{equation} Following the same steps as adopted in the previous
subsection, differentiating Eq. (\ref{OmegaL2}) and using Eqs.
(\ref{Lnonflat}) and (\ref{Ldot}) we obtain
\begin{equation}\label{Hdotnf0}
\frac{\Omega^\prime_d}{\Omega^2_d}=\frac{2}{\Omega_d} \left(
-1+\frac{n'}{n}-\frac{\dot H}{H^2}+\frac{\sqrt{\Omega_d}}{n}
\,\mbox{cosn}(\sqrt{|k|}y) \right).\end{equation} From Friedmann
equation (\ref{FR1nf}) we get
\begin{equation}\label{Hdotnf}
2\frac{\dot H}{H^2}=-3-\Omega_k+\Omega_d+2\frac{\Omega_d^{3/2}}{n}\,
\mbox{cosn}\left(\frac{\sqrt{|k|}R_h}{a}\right)
+2\Omega_d\frac{n^\prime}{n},
\end{equation}
where $\Omega_k\equiv\frac{k}{(aH)^2}$ is the curvature density
parameter. Using Eq. (\ref{Hdotnf}) into Eq. (\ref{Hdotnf0}) we have
{\begin{equation} \label{Omprimenf}
\Omega_d^\prime=\Omega_d\left[1+\Omega_k-\Omega_d+\frac{2\sqrt{\Omega_d}}{n}\,
\mbox{cosn}\left(\frac{\sqrt{|k|}R_h}{a}\right)(1-\Omega_d)\right]
+2\Omega_d(1-\Omega_d)\frac{n^\prime}{n}.
\end{equation}} Here the correction made to the HDE differential equation in
non-flat universe because of the variable $n$ can be observed.
Clearly, when $k=0$ (and thus $\Omega_k=0$) we get Eq.
(\ref{OmegaLdif3}).

\section{Equation of State Parameter ($w(z)$)}
\label{discussion}

We find $w(z)$ for small values of redshifts $z$. Since $\rho_d\sim
a^{-3(1+w)}$, taking the derivatives at the present time $a_0=1$ (so
$\Omega_d=\Omega_d^0$) we get
\begin{equation}
\ln\rho_d =\ln \rho^0_d+{d\ln\rho_d \over d\ln a} \ln a +\frac{1}{2}
{d^2\ln\rho_d \over d(\ln a)^2}(\ln a)^2+\dots....
\end{equation}
Then, $w(\ln a)$ up to second order is given by
\begin{equation}
w(\ln a)=-1-{1\over 3}\left[{d\ln\rho_d \over d\ln a} +\frac{1}{2}
{d^2\ln\rho_d \over d(\ln a)^2}\ln a\right].
\end{equation}
Using $\ln a=-\ln(1+z)\simeq -z$, which is applicable for small
redshifts, one can easily compute $w(z)$, as
\begin{equation}
w(z)=-1-{1\over 3}\left({d\ln\rho_d \over d\ln a}\right)+
\frac{1}{6} \left[{d^2\ln\rho_d \over d(\ln a)^2}\right]\,z\equiv
w_0+w_1z.
\end{equation}

\subsection{Flat FRW geometry}
Using the expression for $\Omega_d'$, given in Eq.
(\ref{OmegaLdif3}) and aforementioned procedure leads to
\begin{eqnarray}
 &&w_0=-{1\over 3}-{2\over 3n}\sqrt{\Omega^0_d}
-\frac{2\Delta_n}{3}\label{w0fl},\\
\label{w1fl}
 &&w_1={1\over
6n}\sqrt{\Omega^0_d}(1-\Omega^0_d)\left(1+{2\over
n}\sqrt{\Omega^0_d}\right)+2\frac{(1-\Omega^0_d)\sqrt{\Omega^0_d}}{6n}\Delta_n.
\ \ \ \ \ \ \ \ \ \ \ \
\end{eqnarray}
These $w_0$ and $w_1$ are for the HDE with varying $n^2$, in a flat
universe. It is clear that when $n$-variation $\Delta_n=0$, we
obtain the results which are consistent with those of
\cite{Li:2004rb}.

The best fit value for $n$ obtained from supernovae type Ia
observational data, within $1-\sigma$ error range \cite{obs3a}, is
$n=0.21$ and from the analysis of $X-$ray gas mass fraction of
galaxy clusters it comes out as $n=0.61$ \cite{obs2}. While combing
the results from different sources we have: the data obtained by the
observations of type Ia supernovae, Cosmic Microwave Background
(CMB) radiation and large scale structure gives $n=0.91$
\cite{obs1}, whereas combining the observations of Baryon Acoustic
Oscillation, $X-$ray gas and type Ia supernovae lead to $n=0.73$
\cite{Wu:2007fs}.

\subsection{Non-flat FRW geometry}
Using the expression of $\Omega_d'$ for non-flat case, given in Eq.
(\ref{Omprimenf}), we have
\begin{eqnarray}\label{w0nonfl}
&&w_0=-{1\over 3}-{2\over
3n}\sqrt{\Omega^0_d}\,\mbox{cosn}\frac{\sqrt{|k|}R_{h0}}{a_0}
-\frac{2 \Delta_n}{3}\\
\label{w1nonfl}
&&w_1=\frac{\sqrt{\Omega^0_d}}{6n}\left[1+\Omega_k^0-\Omega^0_d+\frac{2\sqrt{\Omega^0_d}}{n}\,
\mbox{cosn}\left(\frac{\sqrt{|k|}R_{h0}}{a_0}\right)(1-\Omega^0_d)
\right]\mbox{cosn}\left(\frac{\sqrt{|k|}R_{h0}}{a_0}\right)\nonumber\\&&
+\frac{\Omega^0_d}{3n^2}\,q\left(\frac{\sqrt{|k|}R_{h0}}{a_0}\right)
+2\frac{\sqrt{\Omega^0_d}}{6n}(1-\Omega^0_d)\mbox{cosn}\left(\frac{\sqrt{|k|}R_{h0}}{a_0}\right)\Delta_n,
\end{eqnarray}
where
\begin{equation}q(\sqrt{|k|}y)=\cases{
 \sin^2 y  & \, \, $k=+1$,\\
             0 & \, \, $k=0$,\\
             -\sinh^2 y & \, \, $k=-1$.\\
}\end{equation} Clearly, for $k=0$, Eqs. (\ref{w0nonfl}) and
(\ref{w1nonfl}) reduce to Eqs. (\ref{w0fl}) and (\ref{w1fl})
respectively.

The expressions given by Eqs. (\ref{w0nonfl}) and (\ref{w1nonfl})
involve present values of the parameters $\Omega_d^0$, $\Omega_k^0$,
$a_0$, and $R_{h0}$. From Eq. (\ref{OmegaL2}) we obtain
$L_0=n/(H_0\sqrt{\Omega_\Lambda^0})$. Also from Eq.
(\ref{Lnonflat}), we get
$R_{h0}/a_0=\frac{1}{\sqrt{|k|}}\mbox{sinn}^{-1}(\sqrt{|k|}L_0/a_0)$.
Hence,
\begin{eqnarray}
\frac{R_{h0}}{a_0}&=&\frac{1}{\sqrt{|k|}}\mbox{sinn}^{-1}\left(\frac{n\sqrt{|k|}}{a_0H_0\sqrt{\Omega_\Lambda^0}}\right),
\nonumber\\
&=&
\frac{1}{\sqrt{|k|}}\mbox{sinn}^{-1}\left(\frac{n\sqrt{|\Omega_k^0|}}{\sqrt{\Omega_\Lambda^0}}\right).\label{eqn27}\end{eqnarray}
Substituting Eq. (\ref{eqn27}) in Eqs. (\ref{w0nonfl}) and
(\ref{w1nonfl}), we finally obtain
\begin{eqnarray}\label{w0nonflb}
w_0&=&-{1\over 3}-{2\over 3n}\sqrt{\Omega^0_d-n^2\Omega^0_k}
-2\frac{\Delta_n}{3},\\ \label{w1nonflb}
w_1&=&\frac{\Omega_k^0}{3}+\frac{1}{6n}\sqrt{\Omega^0_d-n^2\Omega_k^0}
\left[1+\Omega_k^0-\Omega^0_d +
\frac{2}{n}(1-\Omega^0_d)\sqrt{\Omega^0_d-n^2\Omega_k^0}\right]\nonumber\\&&
+\frac{1}{6n}\sqrt{\Omega^0_d-n^2\Omega^0_k}
\left(1-\Omega^0_d\right)\Delta_n.
\end{eqnarray}
These $w_0$ and $w_1$ are for non-flat universe, depending only on
$\Omega_d^0$, $\Omega_k^0$, $n$, and $\Delta_n$.

\section{Numerical Results}\label{Numerical Results}
\begin{figure}
\includegraphics [width=8cm]{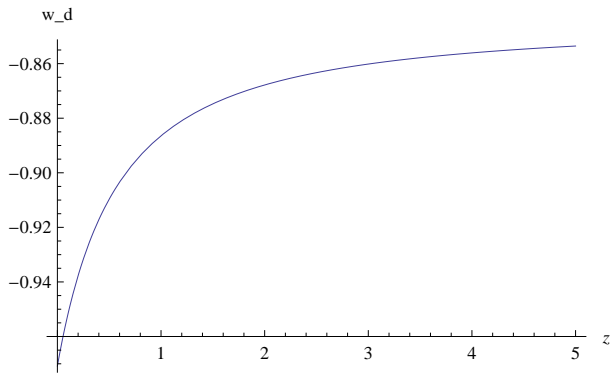}\caption{Evolution of EoS parameter for flat universe versus redshift
parameter $z$. Model parameters $n_0$ and $n_1$ are $0.75$ and $0.1$
respectively. The values for $\Omega_d$ and $\Omega_m$ are taken as
$0.7$ and $0.3$ respectively.} \label{flat}\end{figure}

\begin{figure}
\includegraphics[width=8cm]{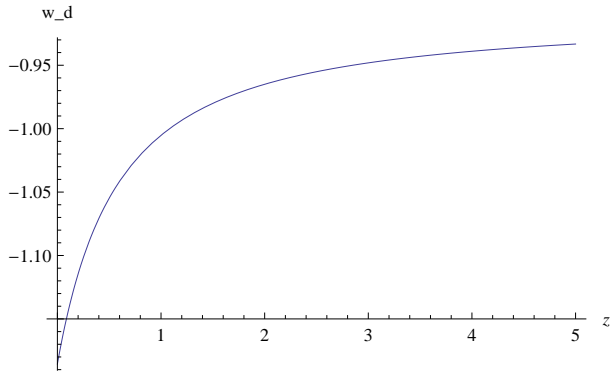}\caption{Evolution of EoS parameter for non-flat universe versus redshift
parameter $z$. Model parameters $n_0$ and $n_1$ are $0.5$ and $0.1$.
The values for $\Omega_d$, $\Omega_m$ and $\Omega_k$ are taken as
$0.7$, $0.285$, and $0.015$ respectively.}
\label{nonflatc00}\end{figure}

By solving the equations for EoS parameters we can give a numerical
description of the evolution of GHDE model. For the choice of model
parameter, $n(z)$, we use the parameterization known as
Chavallier-Polarski-Linder (CPL) \cite{CPL} given as
\begin{equation}n(z)=n_0+n_1\frac{z}{1+z}.\label{CPL}\end{equation}
When $z\rightarrow\infty$ (in the early universe), we see that
$n\rightarrow n_0+n_1$ and as $z\rightarrow0$ (at the present time),
$n\rightarrow n_0$. Therefore, the value of $n$ varies from
$n_0+n_1$ to $n_0$ with passage of time. Also the positive energy
condition of GHDE model requires that
\begin{equation}
n_0>0,~~~~n_0+n_1>0\label{CPL1}.
\end{equation}
Since $w(\ln a)\equiv w_0+w_1z$, using Eq. (\ref{CPL}) and solving
$w_0$ and $w_1$ given in Eqs. (\ref{w0fl}) and (\ref{w1fl}) we plot
the evolutionary behavior of EoS parameter of GHDE model versus
redshift variable (Fig. (\ref{flat})). We choose the values for
$n_0$ and $n_1$ such that they satisfy Eq. (\ref{CPL1}).
 Similarly we
can draw the EoS parameter for non-flat case by solving Eqs.
(\ref{w0nonflb}) and (\ref{w1nonflb}) with Eq. (\ref{CPL}). The
behavior of curve is shown in Fig. (\ref{nonflatc00}). We see that
the curve for EoS parameter enters from $w_d>-1$ to $w_d<-1$. So it
crosses the phantom line, $w_d=-1$, without taking into account the
interaction of dark energy and dark matter.

\section{Conclusions}
\label{Conclusions} In this paper, we have considered the
generalized holographic dark energy model for spatially flat and
non-flat universes with a future event horizon. Since the
holographic parameter, $n$, is generally not constant and can be
assumed as a function of cosmic redshift, we have provided the
complete expressions for cosmological parameters, introducing the
correction terms due to varying $n$. For GHDE with future event
horizon, we have obtained the EoS parameter at small redshifts by
performing the Taylor series expansion up to the first order i.e.
$w(z) \equiv w_0+w_1z$. We have obtained $w_0$ and $w_1$ in terms of
$\Omega_d^0$, $\Omega_k^0$ and $n$-variation $\Delta n$. The
expressions for evolution of the dark energy density have an
additional term depending on $\Delta n$.

For further exploration we have also considered the CPL
parametrization in which $ n(z)=n_0+n_1\frac{z}{1+z}$ . An
investigation has been made for the effect of correction term on EoS
parameters for flat and non-flat geometries. Obtaining numerical
values of the cosmological parameters, we have plotted them against
$z(t)$. Graphical behavior of EoS parameters for our model shows
that with a varying $n$ a shifting from quintessence regime
($w_d>-1$) to phantom regime ($w_d<-1$) is possible, without taking
into account the interaction of dark matter and dark energy. Hence
the GHDE model has a correspondence with the observational data.

\end{document}